# Infrared spectra of $(CO_2)_2$ - Rg trimers, Rg = Ne, Ar, Kr, and Xe

A.J. Barclay,[1] A.R.W. McKellar,[2] and N. Moazzen-Ahmadi[1]

[1] *Department of Physics and Astronomy, University of Calgary, 2500 University Drive North West, Calgary, Alberta T2N 1N4, Canada*

[2] *National Research Council of Canada, Ottawa, Ontario K1A 0R6, Canada*

**Abstract**

High resolution spectra of $(CO_2)_2$-Rg trimers (Rg = Ne, Ar, Kr, and Xe) in the region of the $CO_2$ $\nu_3$ fundamental ($\approx 2350$ cm$^{-1}$) are reported, using a tunable OPO laser source to probe a pulsed supersonic slit jet expansion. These $(CO_2)_2$-Rg transitions tend to be hidden among stronger spectra due to other species, such as $CO_2$-Rg and $(CO_2)_2$. Each trimer consists of a $(CO_2)_2$ unit which is similar to the free carbon dioxide dimer (planar parallel staggered) plus an Rg atom located out-of-plane on the dimer $C_2$ symmetry axis, but the $(CO_2)_2$ unit may not remain exactly planar in the dimer. Experimental structures show that the C-Rg bond lengths in the trimers are similar to those in the corresponding $CO_2$-Rg dimers. As well, the vibrational band origin shifts, relative to $(CO_2)_2$ itself, are similar to those of $CO_2$-Rg relative to $CO_2$.

# 1. Introduction

Weakly bound dimers composed of a carbon dioxide molecule and a rare gas atom have been extensively studied by high resolution spectroscopy. In addition, the trimer $CO_2$-$Ar_2$ is known, and other $CO_2$-$(Rg)_n$ clusters are being studied in our lab. Here, we report on another family of trimers, namely those containing two $CO_2$ molecules and one rare gas atom. The present work was inspired by our recent studies of the related trimers $(CO_2)_2$-CO and $(CO_2)_2$-$N_2$,[1,2] whose structures can be described as resembling a $CO_2$ dimer (near planar, staggered) with the CO or $N_2$ aligned along the $CO_2$ dimer $C_2$ symmetry axis. Not surprisingly, it turns out that $(CO_2)_2$-Rg trimers have similar structures with the Rg atom located on the symmetry axis, as shown in Fig. 1.

High resolution infrared[3-15] and microwave[11,16-20] spectroscopy of $CO_2$-Rg dimers has revealed basic characteristics which are relevant for the trimers studied here. These include the intermolecular separations (C-Rg bond lengths) which are approximately (3.29, 3.50, 3.62, 3.81 Å) for the series (Ne, Ar, Kr, Xe), as well as the vibrational frequency shifts (relative to the free $CO_2$ molecule) for the $CO_2$ $\nu_3$ band which are (+0.130, -0.470, -0.884, -1.471 cm$^{-1}$). These dimer parameters are helpful for estimating expected structures and band origins for the $(CO_2)_2$-Rg trimers. Also relevant here is the spectrum of the $(CO_2)_2$ dimer,[21-26] which has two fundamental vibrations arising from the $CO_2$ monomer $\nu_3$ (asymmetric stretch) mode. Because of the $C_{2h}$ symmetry of the dimer, one of these modes is infrared-inactive while the other active one is responsible for the known $(CO_2)_2$ $\nu_3$ region fundamental band, centered at 2350.771 cm$^{-1}$. This latter band corresponds closely to the $(CO_2)_2$-Rg spectra reported here.

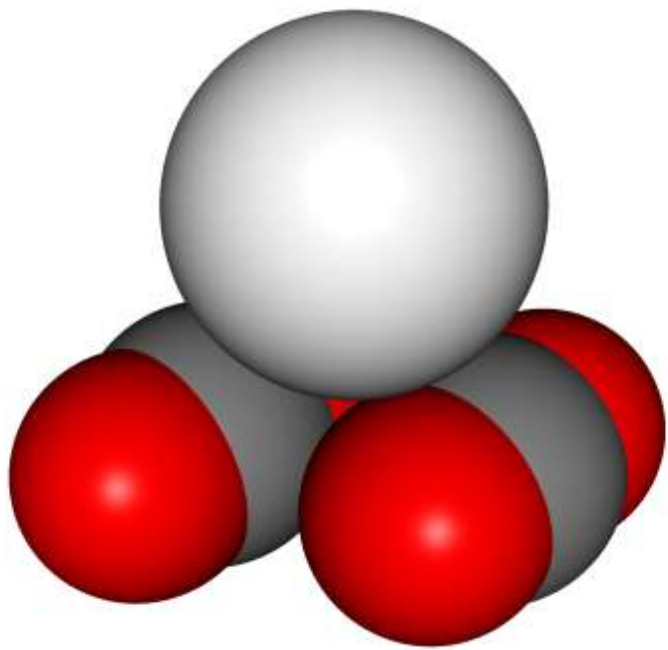

Figure 1: Schematic view of $(CO_2)_2$-Xe. The structure of the $(CO_2)_2$-Rg trimers resembles that of $(CO_2)_2$ (planar parallel staggered) with the Rg located on the $C_2$ symmetry axis at a distance of 2.9 to 3.5 Å from the dimer center of mass.

## 2. Results

The spectra were recorded as described previously,[9,27] using a pulsed supersonic slit jet expansion probed by a rapid-scan optical parametric oscillator source. A typical gas expansion mixture contained about 0.04% $CO_2$ plus 0.5 to 1.0% Ne, Ar, Kr, or Xe in helium carrier gas with a backing pressure of about 13 atmospheres. Wavenumber calibration was carried out by simultaneously recording signals from a fixed etalon and a reference gas cell containing room temperature $CO_2$. Spectral simulation and fitting were made using the PGOPHER software.[28]

It turns out that the $C_2$ symmetry axis which runs through the Rg atom in $(CO_2)_2$-Rg trimers (see Fig. 1) coincides with the $b$-inertial axis for Ne and Ar, but the $a$-axis for Kr and Xe. This has consequences for the rotational selection rules in our observed spectra, as well as nuclear spin statistics (arising from equivalent $CO_2$ molecules and zero nuclear spins for $^{12}C$ and $^{16}O$). As a result, $(CO_2)_2$-Ne and -Ar have $a$- and $c$-type transitions and allowed ground state rotational levels with $(K_a, K_c)$ = (even, even) and (odd, odd), while $(CO_2)_2$-Kr and -Xe have $b$- and $c$-type transitions and allowed ground state levels with $(K_a, K_c)$ = (even, even) and (even, odd). Here the *c*-type transitions are somewhat stronger, with the *a*- or *b*-type ones being roughly 20% weaker, as in the case of $(CO_2)_2$.

The $(CO_2)_2$-Rg spectra are difficult to observe because they tend to be obscured among stronger transitions due to $CO_2$-Rg dimers (including $CO_2$-He, from the He carrier gas), $CO_2$ monomers, $(CO_2)_2$ dimers, and other larger clusters. This probably explains why they have not been reported previously. But we were encouraged by our previous detection[1] of the analogous $(CO_2)_2$-CO trimer, and by the fact that it was straightforward to estimate the $(CO_2)_2$-Rg structures (and hence rotational parameters) with reasonable accuracy. The strongest and most obvious feature in the $(CO_2)_2$-Rg spectra is a central *Q*-branch peak. Figure 2 shows the example of $(CO_2)_2$-Xe, for which this *Q*-branch appears at 2349.24 cm$^{-1}$ in a relatively uncrowded region of the spectrum. It is evident that all the other spectral peaks of $(CO_2)_2$-Xe, in the *P*- and *R*-branch regions, are much weaker than the *Q*-branch.

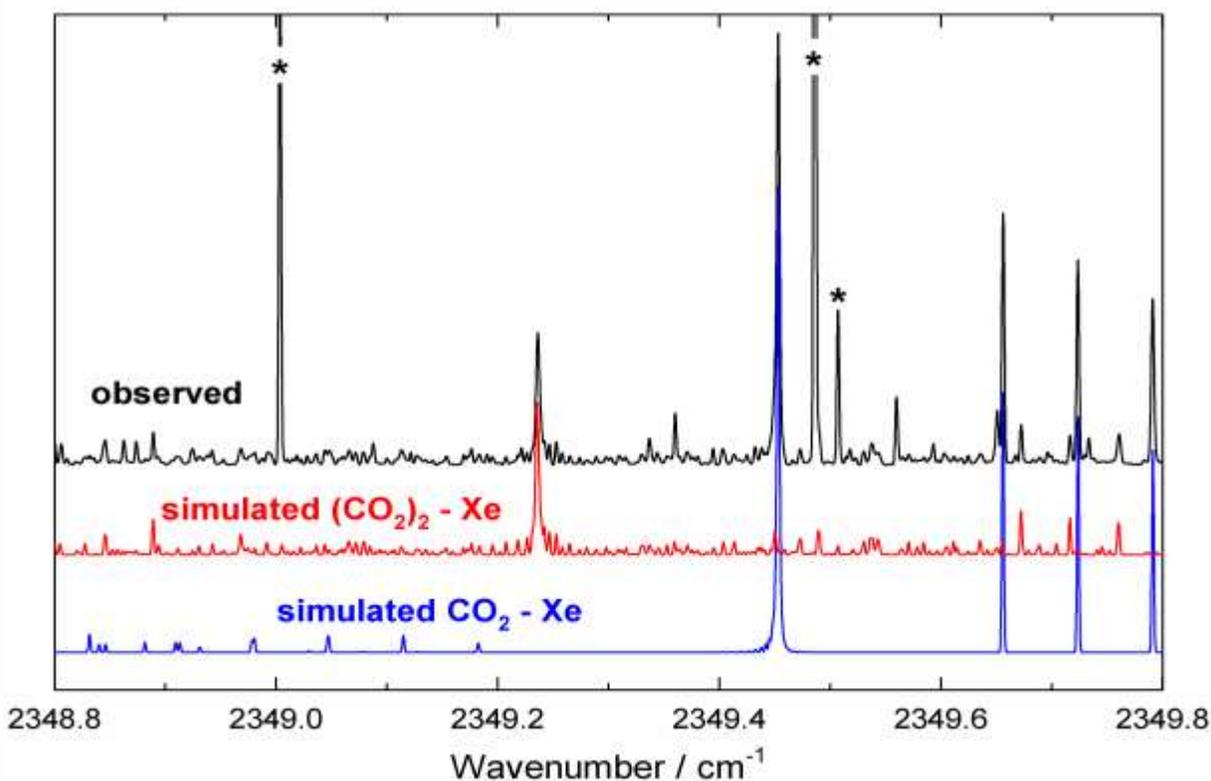

Figure 2: Observed and simulated (T=2K) spectra of $(CO_2)_2$-Xe, also showing $CO_2$-Xe.[15] The strongest feature in the $(CO_2)_2$-Xe spectrum is the *Q*-branch at 2349.24 cm$^{-1}$. Lines marked with an asterisk are due to $CO_2$-He.

The pattern of the present $(CO_2)_2$-Rg spectra is illustrated in Fig. 3 by simulated spectra, and the parameters resulting from fits to the spectra are listed in Table 1. The most difficult case was that of $(CO_2)_2$-Kr, in part because its $Q$-branch happens to be obscured by $CO_2$ monomer absorption ($R(0)$ of the fundamental band). This is reflected in larger standard deviations for the $(CO_2)_2$-Kr parameters in Table 1. A notable feature of these parameters is that the $B$-values of $(CO_2)_2$-Ne and -Ar and the $A$-values of $(CO_2)_2$-Kr and -Xe are nearly equal to each other ($\approx 0.0463$ cm$^{-1}$) and in turn very similar to the $C$-value of $(CO_2)_2$ itself (0.0453 cm$^{-1}$). This is just what we expect from the fact that the $(CO_2)_2$-Rg trimers resemble $(CO_2)_2$ with an Rg atom added on the symmetry axis.

In the case of both $(CO_2)_2$-CO and -$N_2$, we observed two fundamental bands in the $CO_2$ $\nu_3$ region, a stronger ($a,c$)-type band corresponding to the $(CO_2)_2$-Rg bands reported here, and a weak $b$-type band.[2] Detection of the weak band confirmed that the $(CO_2)_2$ dimer subunits in $(CO_2)_2$-CO and -$N_2$ are not planar, and calculations indicated that each $CO_2$ unit departs from planarity by about 19°. We searched for the analogous second fundamentals for $(CO_2)_2$-Rg, but were not able to observe them. This does not prove exact planarity, but it probably indicates that the $(CO_2)_2$ dimer subunits are more nearly planar for $(CO_2)_2$-Rg than for $(CO_2)_2$-CO and -$N_2$.

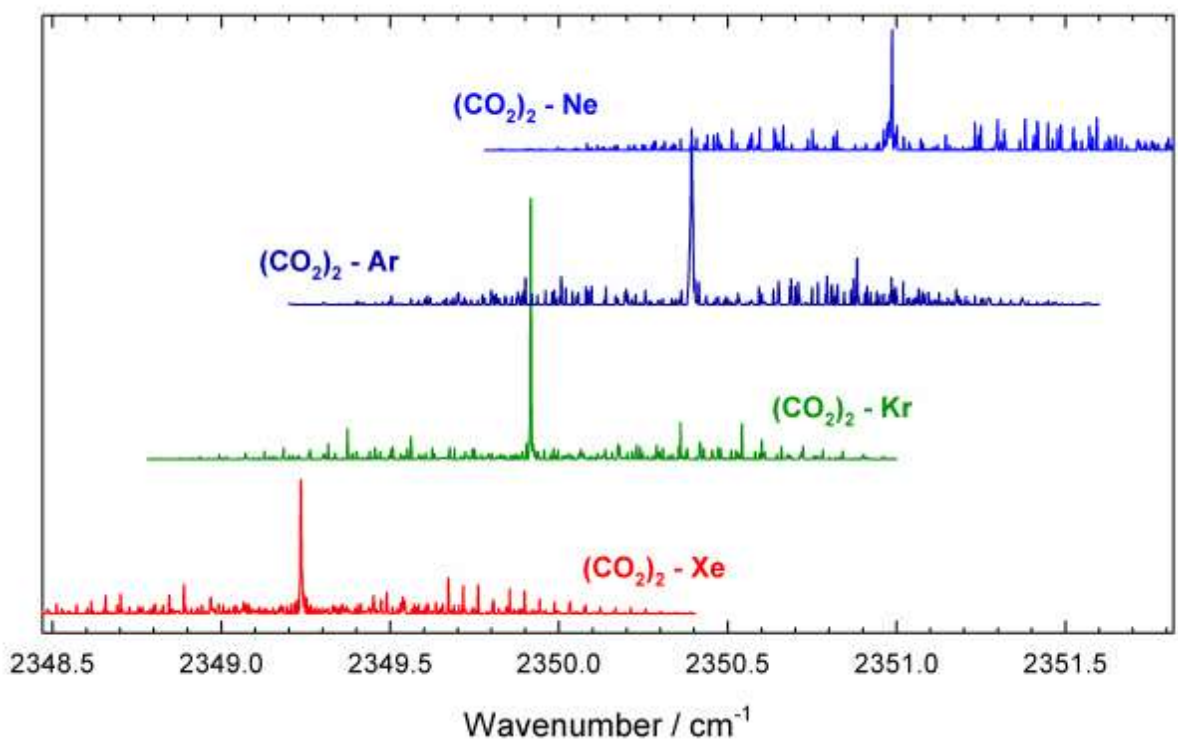

Figure 3: Simulated (T=2K) spectra of the $(CO_2)_2$-Rg trimers studied here.

Table 1. Molecular parameters for $(CO_2)_2$-Rg trimers (in cm$^{-1}$). [a]

|  | $(CO_2)_2$-Ne | $(CO_2)_2$-Ar | $(CO_2)_2$-Kr | $(CO_2)_2$-Xe |
| --- | --- | --- | --- | --- |
| $\nu_0$ | 2350.9886(1) | 2350.3958(1) | 2349.9177(3) | 2349.2372(1) |
| $A'$ | 0.086862(86) | 0.053036(10) | 0.046222(21) | 0.046186(11) |
| $B'$ | 0.046072(19) | 0.046248(13) | 0.033951(69) | 0.025311(14) |
| $C'$ | 0.037134(18) | 0.0294297(71) | 0.022621(82) | 0.018236(22) |
| $A''$ | 0.086943(78) | 0.052984(10) | 0.046404(23) | 0.046354(13) |
| $B''$ | 0.046240(20) | 0.046424(14) | 0.033905(85) | 0.025265(12) |
| $C''$ | 0.037307(24) | 0.0294960(72) | 0.022608(97) | 0.018254(21) |
| $10^5 \times \Delta_{JK}$ | 1.07(24) |  |  |  |
| $n$ | 26 | 52 | 24 | 45 |
| rmsd | 0.00027 | 0.00031 | 0.00096 | 0.00034 |

[a] Quantities in parentheses are 1σ from the least-squares fit, in units of the last quoted digit. $n$ is the number of observed lines and rmsd is the root mean square average error in the fit.

## 3. Discussion and conclusions

### 3.1. Structures

Assuming $C_2$ symmetry, four structural parameters are required to specify the structure of a $(CO_2)_2$-Rg trimer. These can be specified as: $R_1$ (center of mass (c.m.) separation of Rg and $(CO_2)_2$ subunits); $R_2$ (c.m. separation of $CO_2$ subunits, that is, C-C distance); $\theta$ (angle between a line connecting the C atoms of the $CO_2$ units and an OCO axis); and $\phi$ (departure from planarity of $(CO_2)_2$ subunit). But we only have three rotational constants for each trimer, and so cannot determine unique experimental structures. In order to obtain structures here, we assume $\phi = 0$ (planar $(CO_2)_2$ subunit) and this yields the structural parameters listed in Table 2 (note that those for $(CO_2)_2$-Kr are probably less reliable than the others, as mentioned above). For comparison, analogous parameters for $(CO_2)_2$ and $CO_2$-Rg dimers are included. The values of $R_2$ (3.55 – 3.60 Å) and $\theta$ (56 - 59°) are similar for all the $(CO_2)_2$-Rg trimers and for $(CO_2)_2$. It appears that the presence of the nearby Rg atom causes the $CO_2$ monomers to move slightly closer together. The values of $R_1$ naturally increase with the Rg mass, as do the resulting values of $R(C-Rg)$ which are similar to those of the corresponding $CO_2$-Rg dimers.

Table 2. Experimental structures for $(CO_2)_2$-Rg trimers, assuming planar $(CO_2)_2$ subunit (in Å, or ° for $\theta$).[a]

|  | $\theta$ | $R_2$ | $R_1$ | $R(C-Rg)$ trimer | $R(C-Rg)$ dimer [10] |
|---|---|---|---|---|---|
| $(CO_2)_2$ [25] | 57.9 | 3.60 |  |  |  |
| $(CO_2)_2$-Ne | 56.2 | 3.56 | 2.94 | 3.43 | 3.29 |
| $(CO_2)_2$-Ar | 57.3 | 3.55 | 3.10 | 3.57 | 3.50 |
| $(CO_2)_2$-Kr | 59.0 | 3.55 | 3.20 | 3.66 | 3.62 |
| $(CO_2)_2$-Xe | 56.5 | 3.55 | 3.41 | 3.85 | 3.81 |

[a] $\theta$ is angle between the C-C and OCO axes; $R_2$ is the C-C distance; $R_1$ is the Rg to $(CO_2)_2$ center of mass distance; and $R(C-Rg)$ is the Rg to C atom distance for the trimer and (for comparison) the corresponding dimer.

## 3.2. Vibrational shifts

The shifts of the $(CO_2)_2$-Rg band origins with respect to that of $(CO_2)_2$ are listed in Table 3. There is a systematic trend from a small blue shift for $(CO_2)_2$-Ne to a much larger red shift for $(CO_2)_2$-Xe. Interestingly, this trend is remarkably similar to that determined previously for $CO_2$-Rg dimers,[25] as illustrated in two ways in Fig. 4. The lower panel in Fig. 4 also shows how both the dimer and trimer shifts have an almost linear dependence on the polarizability of the rare gas atom. The relatively small differences between the dimer and trimer shifts may be of theoretical interest for future modeling of potential surfaces and nonadditive three body effects.

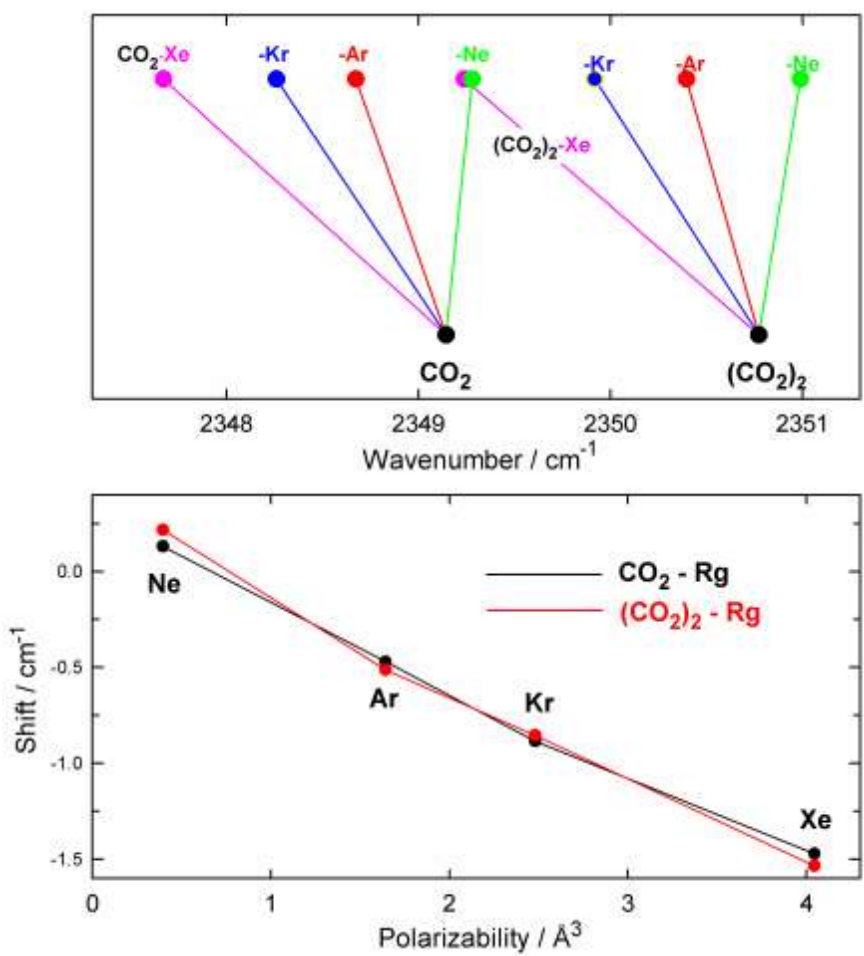

Figure 4: Upper panel: schematic view of vibrational shifts for $CO_2$-Rg dimers relative to $CO_2$ monomer, compared with $(CO_2)_2$-Rg trimers relative to $(CO_2)_2$ dimer. Lower panel: dependence of the vibrational shifts of $CO_2$-Rg dimers and $(CO_2)_2$-Rg trimers on the polarizability of the rare gas atom.

Table 3. Vibrational shifts for $(CO_2)_2$-Rg trimers, with respect to $(CO_2)_2$, compared to those for $CO_2$-Rg dimers, with respect to $CO_2$ (in cm$^{-1}$).

| Rg | $(CO_2)_2$-Rg | $CO_2$-Rg [10] |
|----|---------------|----------------|
| Ne | +0.217        | +0.130         |
| Ar | -0.513        | -0.470         |
| Kr | -0.854        | -0.884         |
| Xe | -1.534        | -1.471         |

### 3.3. Conclusions

It would be very interesting to observe the remaining trimer in the present family, $(CO_2)_2$-He, but so far we have not been able to find it. Also interesting would be the tetramers $(CO_2)_2$-$(Rg)_2$. Would their structures involve Rg atoms on the symmetry axis on either side of the $(CO_2)_2$, or would the Rg atoms locate on one side, forming an $(Rg)_2$ dimer? This could change, depending on the Rg atom. Mixed tetramers involving two different Rg atoms are also possible.

Another avenue for future investigation involves combination bands of the present trimers involving excitation of the $CO_2$ $\nu_3$ mode plus a low frequency intermolecular mode. Such bands have already been observed[2] for both $(CO_2)_2$-CO and -$N_2$. They are relatively weak, but have the advantage of being more likely to appear in spectral regions with less interference from other species.

In conclusion, weakly-bound trimers of the form $(CO_2)_2$-Rg (Rg = Ne, Ar, Kr, Xe) have been observed by high-resolution infrared spectroscopy in the region of the $CO_2$ $\nu_3$ fundamental band ($\approx$2350 cm$^{-1}$). Their structures resemble the carbon dioxide dimer, $(CO_2)_2$ (near planar, side-by-side staggered) with an Rg atom located on its out-of-plane $C_2$ symmetry axis. Experimental structures derived from the observed rotational constants (assuming planarity for the $(CO_2)_2$), confirm the similarity of the $(CO_2)_2$ unit in the trimers to the carbon dioxide dimer and give C-Rg distances similar to those in the corresponding $CO_2$-Rg dimers. Vibrational shifts of the $(CO_2)_2$-Rg trimer band origins relative to $(CO_2)_2$ are found to be remarkably similar to those of $CO_2$-Rg dimers relative to $CO_2$.

10**Acknowledgements**

The financial support of the Natural Sciences and Engineering Research Council of Canada is gratefully acknowledged.